\documentclass[11pt,journal]{IEEEtran}
\usepackage{latexsym}
\usepackage{amsfonts,amssymb,amsmath}
\usepackage{mathtools, cuted}
\usepackage{hyperref}
\usepackage{amsmath}
\usepackage[T1]{fontenc}
\usepackage{cite}
\usepackage{subcaption}
\usepackage{overpic}
\usepackage{array}
\usepackage{url}
\usepackage{color}
\usepackage{float}
\usepackage{units}
\usepackage{comment}
\usepackage{balance}

\newcommand{\cmmnt}[1]{\ignorespaces}

\usepackage[subtle]{savetrees}

\title{From Antenna Abundance to Antenna Intelligence\\ in 6G Gigantic MIMO Systems}

\author{Emil Bj{\"o}rnson, Amna Irshad, \"Ozlem Tu\u{g}fe Demir, Alva Kosasih, Giuseppe Thadeu Freitas de Abreu, Vitaly Petrov 
\thanks{E.~Bj{\"o}rnson, A.~Irshad, and V.~Petrov are with the KTH Royal Institute of Technology, Sweden; \"O.~T.~Demir is with Bilkent University, Turkiye; A.~Kosasih is with Nokia Technology Standards, Finland; G. T. F. de Abreu is with the Constructor University Bremen, Germany. The research was supported by the Grant 2022--04222 from the Swedish Research Council and by the SweWIN center (Vinnova grant 2023-00572).}}

\begin{document}

\maketitle

\begin{abstract}
Current cellular systems achieve high spectral efficiency through Massive MIMO, which leverages an abundance of antennas to create favorable propagation conditions for multiuser spatial multiplexing. Looking towards future networks, the extrapolation of this paradigm leads to systems with many hundreds of antennas per base station, raising concerns regarding hardware complexity, cost, and power consumption. This article suggests more intelligent array designs that reduce the need for excessive antenna numbers. We revisit classical uniform array design principles and explain how their uniform spatial sampling leads to unnecessary redundancies in practical deployment scenarios. By exploiting non-uniform sparse arrays with site-specific antenna placements---based on either pre-optimized irregular arrays or real-time movable antennas---we demonstrate how superior multiuser MIMO performance can be achieved with far fewer antennas. These principles are inspired by previous works on wireless localization. We explain and demonstrate numerically how these concepts can be adapted for communications to improve the average sum rate and similar metrics. The results suggest a paradigm shift for future antenna array design, where antenna intelligence replaces sheer antenna count. This opens new opportunities for efficient, adaptable, and sustainable Gigantic MIMO systems.
\end{abstract}

\vspace{-4mm}

\section*{Introduction}

Cellular networks consist of a vast number of base stations (BSs) deployed to provide wireless services to mobile users whenever and wherever needed. Each BS has a designated geographical coverage region and must utilize a multiple-access scheme to divide the radio resources among its users.
The first four network generations used orthogonal access schemes based on time, frequency, or code division.
By contrast, the fifth generation (5G) switched to spatial-division multiple access (SDMA), also known as multiuser MIMO (multiple-input multiple-output), which differentiates multiple users through adaptive beamforming.
The BS uses an array of $M$ antennas to point distinct beams toward $K$ users; thus, it can communicate with them simultaneously over the entire frequency band.
If the beams are designed based on channel state information (CSI) to optimally balance between signal amplification and co-user interference suppression, SDMA enables a multiplexing gain of $\min(M,K)$. This means that the sum rate (bit/s/Hz) is proportional to the multiplexing gain when the signal-to-noise ratio (SNR) is high and greatly surpasses the sum rate achieved with traditional orthogonal access schemes.

\begin{figure}[t!]
        \centering 
        \begin{subfigure}[b]{\columnwidth} \centering 
	\begin{overpic}[width=\columnwidth,tics=10]{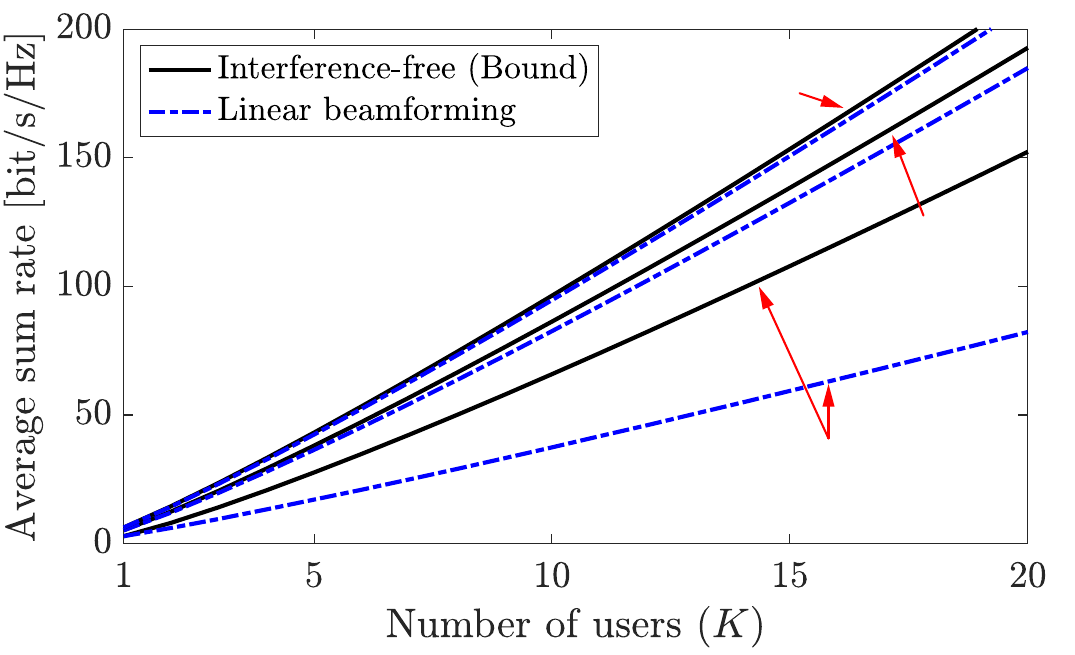} 
 \put(75,16){\small $\frac{M}{K}=1$}
  \put(84.5,36.5){\small $\frac{M}{K}=4$}
   \put(63.5,52){\small $\frac{M}{K}=8$}
 \end{overpic}  
                \caption{Uplink sum rate with different antenna/user ratios, iid Rayleigh fading channels, and $10$\,dB SNR per antenna.}  \vspace{+3mm}
        \end{subfigure} 
        \begin{subfigure}[b]{\columnwidth} \centering 
	\begin{overpic}[width=\columnwidth,tics=10]{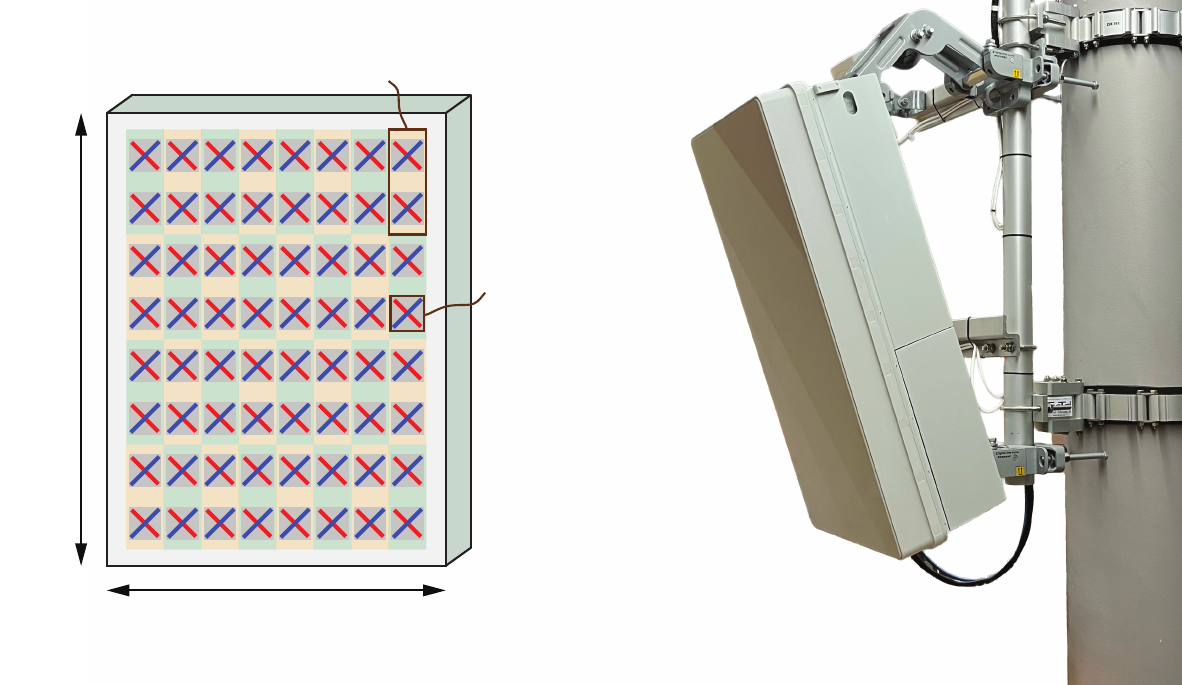} 
                \put(2.5,23){\small \rotatebox{90}{540 cm}}
  \put(17.5,4){\small 370 cm}
  \put(7,52){\small Dual-polarized antenna}
  \put(42,32){\small Radiating}
  \put(42,28){\small  element}
 \end{overpic}  \vspace{-8mm}
                \caption{A typical 5G Massive MIMO antenna array.} 
        \end{subfigure}   
        \caption{Massive MIMO in 5G builds on having an antenna abundance. As shown in (a), the sum rate with RZF beamforming becomes close to the interference-free bound when $M/K \geq 4$. The practical BS shown in (b) has $M=32$ dual-polarized antennas, each comprising two discrete radiating elements (drawn as crosses) to enhance the antenna gain. This BS is designed to support $M/K=4$.}
\label{fig_massivemimo} 
\end{figure}
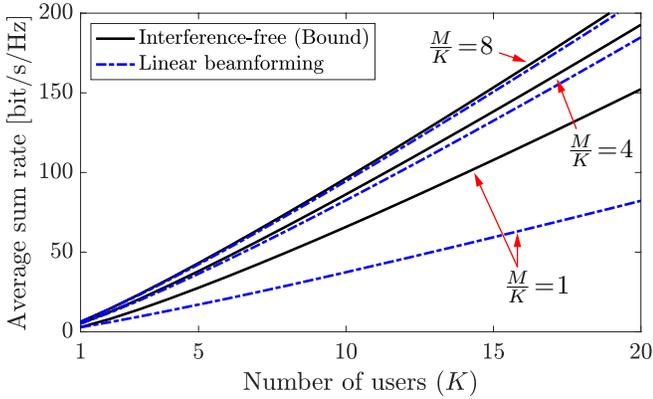
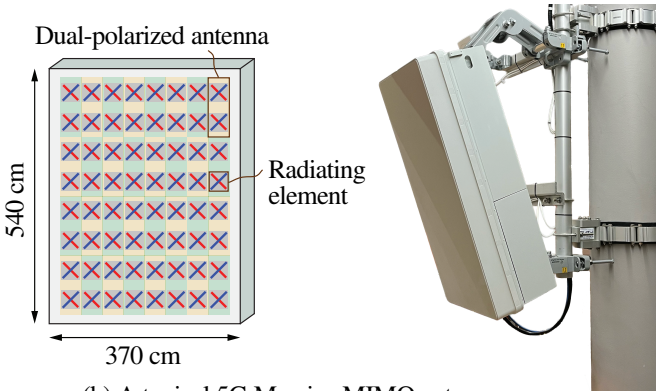

SDMA was conceived as early as the 1990s~\cite{Swales1990a}, but it was then commonly assumed that $M \approx K$ because the multiplexing gain is not improved when one parameter deviates from the other.
In 2010, Marzetta challenged this assumption with his \emph{Massive MIMO} concept where $M \gg K$~\cite{Marzetta2010a}. 
Fig.~\ref{fig_massivemimo}(a) shows simulation results that highlight the main motivation for having this abundance of antennas.
The figure shows how the sum rate grows with the number of users when $M/K$ equals $1$, $4$, or $8$. The sum rate is averaged over many channel realizations, generated using independent and identically distributed (iid) Rayleigh fading.
Dashed curves indicate the performance achieved by regularized zero-forcing (RZF) beamforming based on perfect CSI, while solid curves indicate the theoretical maximum if co-user interference is neglected. There is a significant gap between the solid and dashed curves when $M/K=1$, but the gap narrows when $M/K$ increases (i.e., with a surplus of antennas).
This is attributed to the \emph{favorable propagation} phenomenon, namely, that a set of randomly generated channel vectors is likely to be nearly mutually orthogonal when the vector dimension (number of antennas) is much larger than the number of vectors (number of users).

5G networks may use Massive MIMO in the $3.5$\,GHz band, where $M=32$ dual-polarized antennas fit into a typical BS with the physical dimensions shown in Fig.~\ref{fig_massivemimo}(b). The $\pm 45^\circ$-polarized antennas (illustrated in red and blue, respectively) enable the multiplexing of two data streams per user, as 5G devices also have dual-polarized antennas.
The measurement campaign in~\cite{Signals2020a} demonstrated how a commercial BS delivered $710$\,Mbps when serving one user, while it increased to $5.2$\,Gbps when spatially multiplexing $K=8$ users. 
The average rate was $650$\,Mbps per user, a mere $8.5\%$ reduction compared to the single-user case, indicating that beamforming handled interference effectively. The antenna-user ratio was $M/K=4$ in this 5G setup.

Based on the success of 5G Massive MIMO, it is logical to target even larger multiplexing gains in forthcoming sixth-generation (6G) networks.
The new 6G spectrum will be in the upper mid-band, from $7$--$15$ GHz, where the wavelength is $2$--$4$ times smaller than in the $3.5$\,GHz band~\cite{Bjornson2025a}.
If we keep the uniform planar array (UPA) geometry, shown in Fig.~\ref{fig_massivemimo}(b), and given that antennas are proportionally sized to the wavelength, we can fit $4$--$16$ times more antennas into an equal-sized box and increase the multiplexing gain accordingly while maintaining $M/K \geq 4$.
Such \emph{Gigantic MIMO}~\cite{Bjornson2025a} dimensions require an enormous amount of radio electronics and lead us to the core question behind this paper:

\emph{Can we avoid Massive-MIMO-like antenna abundance in 6G with a more intelligent array design?}

The answer is yes in many practical scenarios. In the remainder of this paper, we will step-by-step explain how one can tailor antenna arrays to network and site-specific needs. We will revisit the impact of antenna spacings in uniform arrays, describe methods for designing non-uniform sparse arrays, investigate the role that movable antennas can play, and conclude by discussing how to link budget can be maintained when having fewer antennas.

\vspace{-2mm}

\section*{Revisiting Uniform Array Designs}

An antenna array can have infinitely many different shapes, but it is customary to use uniform arrangements in BSs. Uniform linear arrays (ULAs) and UPAs with half-wavelength antenna spacing are particularly common and are easily scaled to different antenna numbers.
In this section, we will first elaborate on the background to this practice and then challenge it.

Suppose we can deploy $M$ isotropic BS antennas arbitrarily along a line, possibly on the edge of a rooftop. If a device sends an uplink signal over a far-field line-of-sight (LOS) channel, then each receive antenna captures the same amount of signal and noise power. If we combine the $M$ received signals using maximum ratio combining, we will achieve an SNR proportional to $M$ since the signals are combined coherently while the noise adds non-coherently. The factor $M$ is known as the \emph{array gain} or \emph{beamforming gain}~\cite{bjornson2024introduction}.
Note that no assumption regarding the antenna arrangement was made to reach this result. However, we implicitly assumed the antennas operate independently, i.e., there is no mutual coupling. As a rule-of-thumb, this requires a spacing of \emph{at least} $\lambda/2$ between any antenna pair, where $\lambda$ is the signal's wavelength~\cite{Yuan2023a}. Hence, a minimum-sized ULA with negligible antenna coupling has $\lambda/2$-spacing.

\begin{figure}[t!]
        \centering 
        \begin{subfigure}[b]{\columnwidth} \centering 
	\begin{overpic}[width=\columnwidth,tics=10]{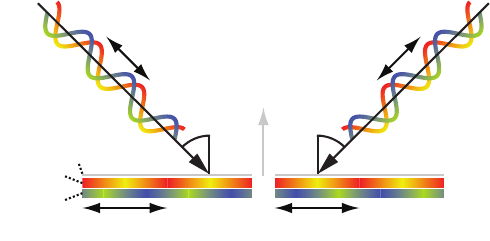}
   \put(48,28.5){\small Normal}
   \put(5,12){\small Real}
   \put(0,5){\small Imaginary}
   \put(0,17.5){\small Observation line}
                \put(26.2,38.6){\small \rotatebox{-45}{$\lambda$}}
            \put(78,37.3){\small \rotatebox{45}{$\lambda$}}
    \put(21,1){\small $\sqrt{2}\lambda$}
    \put(61,1){\small $\sqrt{2}\lambda$}
    \put(39,23.5){\small $\frac{\pi}{4}$}
        \put(63,23.5){\small $-\frac{\pi}{4}$}
 \end{overpic}  
                \caption{The apparent wavelength observed along a line depends on the incident angle, but is generally larger than the original wavelength.}   \vspace{+3mm}
        \end{subfigure} 
        \begin{subfigure}[b]{\columnwidth} \centering
	\begin{overpic}[width=\columnwidth,tics=10]{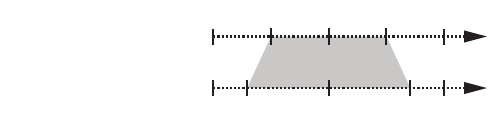}
 
 \put(11,17){\small Incident angle: $\varphi$}
 \put(0,6.5){\small Apparent spatial frequency:}
 \put(10,2.3){\small ($\sin(\varphi)/\lambda$)}
 
  \put(89.5,1){\small $\frac{1}{\lambda}$}
    \put(80.5,1){\small $\frac{1}{\sqrt{2}\lambda}$}
      \put(66.2,1){\small $0$}
        \put(46,1){\small $-\!\frac{1}{\sqrt{2}\lambda}$}
        \put(39,1){\small $-\frac{1}{\lambda}$}
 
   \put(89.5,21.3){\small $\frac{\pi}{2}$}
    \put(77.5,21.3){\small $\frac{\pi}{4}$}
      \put(66.2,21.3){\small $0$}
        \put(51,21.3){\small $-\frac{\pi}{4}$}
 \put(39,21.3){\small $-\frac{\pi}{2}$}
\end{overpic} \vspace{-2mm}
                \caption{Mapping from angles to spatial frequencies in linear arrays.}  \vspace{+3mm}
        \end{subfigure} 
        \begin{subfigure}[b]{\columnwidth} \centering
	\begin{overpic}[width=\columnwidth,tics=10]{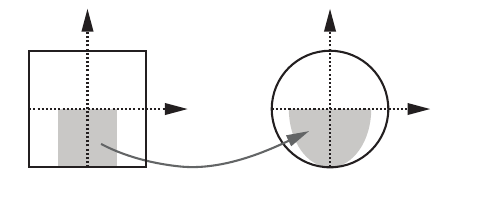}
 \put(70,36.5){\small Vertical}
 \put(70,32.5){\small spatial frequency}
 \put(81,13){\small Horizontal}
 \put(81,9){\small spatial}
 \put(81,5){\small frequency}
 \put(20,36.5){\small Vertical}
 \put(20,32.5){\small angle}
 \put(32,13){\small Horizontal}
 \put(32,9){\small angle}
 \put(30.5,21){\small $\frac{\pi}{2}$}
 \put(0,17.5){\small $-\frac{\pi}{2}$}
 \put(13.5,33){\small $\frac{\pi}{2}$}
 \put(13.5,2.5){\small $-\frac{\pi}{2}$}
 \put(79.5,21){\small $\frac{1}{\lambda}$}
 \put(49,17.5){\small $-\frac{1}{\lambda}$}
 \put(63.5,33){\small $\frac{1}{\lambda}$}
 \put(63,2.5){\small $-\frac{1}{\lambda}$}
\end{overpic} 
                \caption{Mapping from angles to spatial frequencies in planar arrays.}  
        \end{subfigure} 

        \caption{A carrier wave with a fixed wavelength $\lambda$ gives rise to different apparent wavelengths when observed on a line or plane. (a) illustrates how the apparent wavelength varies with the incident angle. (b) shows how the mapping between incident angles and apparent spatial frequencies. The shaded region demonstrates the nonlinearity of the mapping. (c) shows the corresponding mapping for a planar array.}
\label{fig_spatialfreq} 
\end{figure}
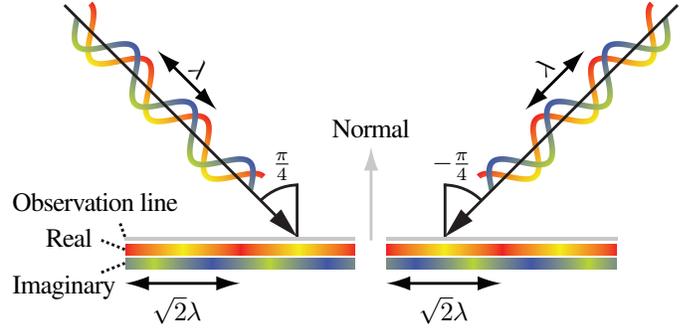
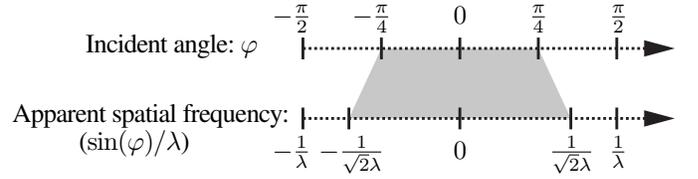
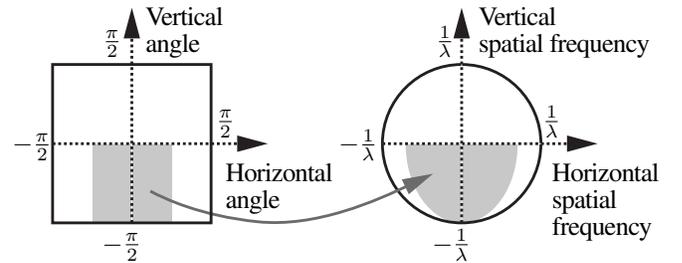

There are other classical ways to motivate this specific design.
Consider a planar carrier wave with wavelength $\lambda$, which is the length of each cycle of the sinusoid (i.e., the distance between two peaks in the real or imaginary part).
When this wave propagates in free space and is observed along a line, the projection stretches the cycle lengths depending on the incident angle.
This fact is illustrated in Fig.~\ref{fig_spatialfreq}(a) for the incident angles $+\pi/4$ and $-\pi/4$, for which $\sin(\pm \pi/4)=\pm 1/\sqrt{2}$. This results in \emph{apparent} wavelengths of $+\sqrt{2}\lambda$ and $-\sqrt{2}\lambda$, respectively, if we measure distances between peak values along the line.
The sign indicates whether the wave's angle-of-arrival is positive or negative (i.e., left or right of the normal of the observation line).
The sign carries over to the observed signal's imaginary part, as shown in the figure, so every angle from $-\pi/2$ to $\pi/2$ is observed with a unique pattern.

As the carrier wave has a spatial cycle length of $\lambda$, its \emph{spatial frequency} is defined as the inverse of the wavelength: $1/\lambda$.
Fig.~\ref{fig_spatialfreq}(b) shows how the \emph{apparent} spatial frequency along the observation line varies between $+1/\lambda$ and $-1/\lambda$, depending on the incident angle.
Since the incident signal can have any spatial frequency within this range, the signal's \emph{spatial bandwidth} is $2/\lambda$. This range is particularly relevant in a multipath propagation environment, where multiple copies of the same signal arrive with different spatial frequencies, and in a multiuser communication scenario where each user is associated with a distinct combination of spatial frequencies.
The Nyquist-Shannon sampling theorem for complex signals \cite[p.~87]{bjornson2024introduction} prescribes that one should sample such signals at points separated by \emph{at most} the inverse of the spatial bandwidth to avoid spatial aliasing (i.e., that signals arriving from different angles give identical samples).
Together with the condition for mutual coupling avoidance, this suggests the use of a ULA with antennas \emph{exactly} $\lambda/2$ apart.

Yet another argument in favor of this array design stems from non-LOS multipath propagation.
In the richest kind of fading environment---called 3D isotropic scattering---the correlation between the channel realizations observed at two locations is a sinc-function with nulls at integer multiples of $\lambda/2$ \cite[p.~321]{bjornson2024introduction}.
Hence, one obtains iid fading with maximum spatial diversity using ULAs with that spacing.

Despite these theoretical arguments, other array geometries are often preferred in practice.
Cellular networks are sectorized, which means that each BS array only covers a limited range of horizontal angles with a width of $60^\circ$--$120^\circ$.
Suppose the actual range is from $-\pi/4$ to $+\pi/4$, as illustrated by the shading in Fig.~\ref{fig_spatialfreq}(b). This shrinks the spatial bandwidth of the incoming signals by a factor of $1/\sqrt{2} \approx 0.71$ and thereby increases the antenna spacing needed for fading decorrelation and prescribed by the sampling theorem by a factor of $\sqrt{2} \approx 1.41$.
Furthermore, many cellular BSs use UPAs that can also distinguish between signals arriving from different vertical elevation angles, which is particularly important in urban deployments with high-rise buildings.
If the array is deployed above the coverage region (e.g., on a rooftop or tower), then signals from user devices can only arrive from below. 
Thus, if all signals arrive within the vertical angular range $-\pi/2$ to $0$, the spatial bandwidth shrinks by a factor of $0.5$, increasing the recommended antenna spacing from $\lambda/2$ to $\lambda$. The shaded region in Fig.~\ref{fig_spatialfreq}(c) illustrates the set of mutually observable horizontal and vertical spatial frequencies in this example.
In conclusion, if the array is designed for a specific deployment scenario rather than the worst-case situation, there are compelling reasons to consider antenna spacings far beyond $\lambda/2$.

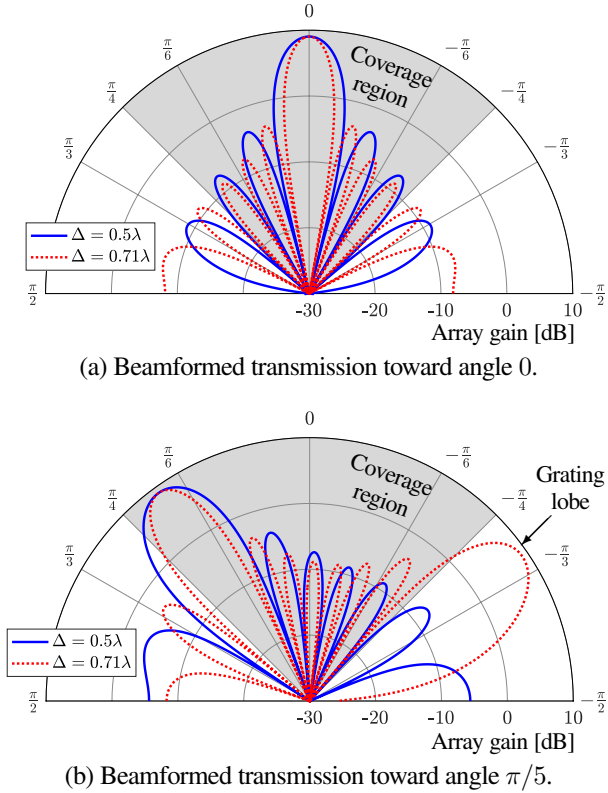
\begin{figure}[t!]
        \centering 
        \begin{subfigure}[b]{\columnwidth} \centering 
	\begin{overpic}[width=.9\columnwidth,tics=10]{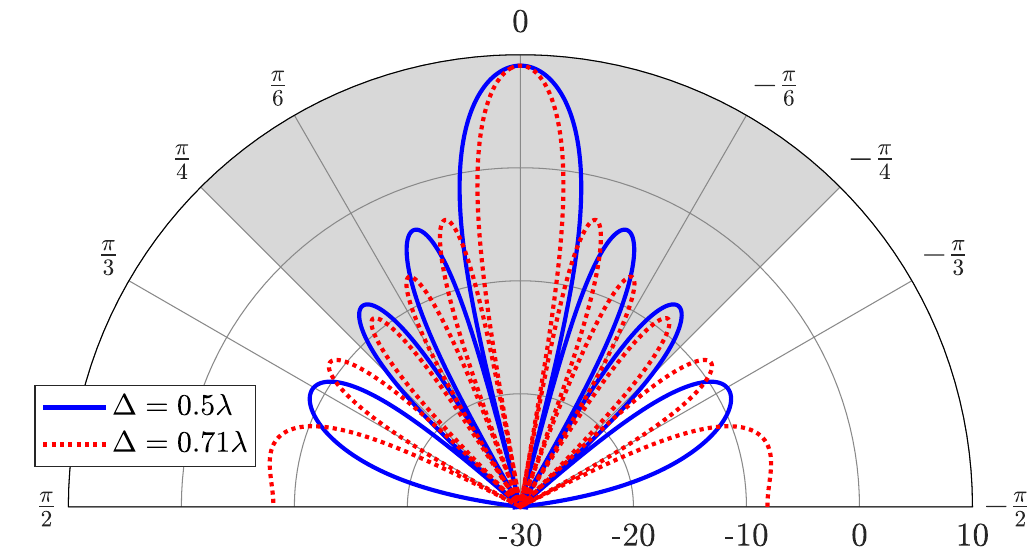} 
	 \put(57,44){\footnotesize \rotatebox{-20}{Coverage}}
	 \put(57.5,39){\footnotesize \rotatebox{-20}{region}}
 \put(71,-3){\footnotesize Array gain [dB]}
 \end{overpic}  \vspace{2mm}
                \caption{Beamformed transmission toward angle $0$.}   \vspace{+3mm}
        \end{subfigure} 
        \begin{subfigure}[b]{\columnwidth} \centering
	\begin{overpic}[width=.9\columnwidth,tics=10]{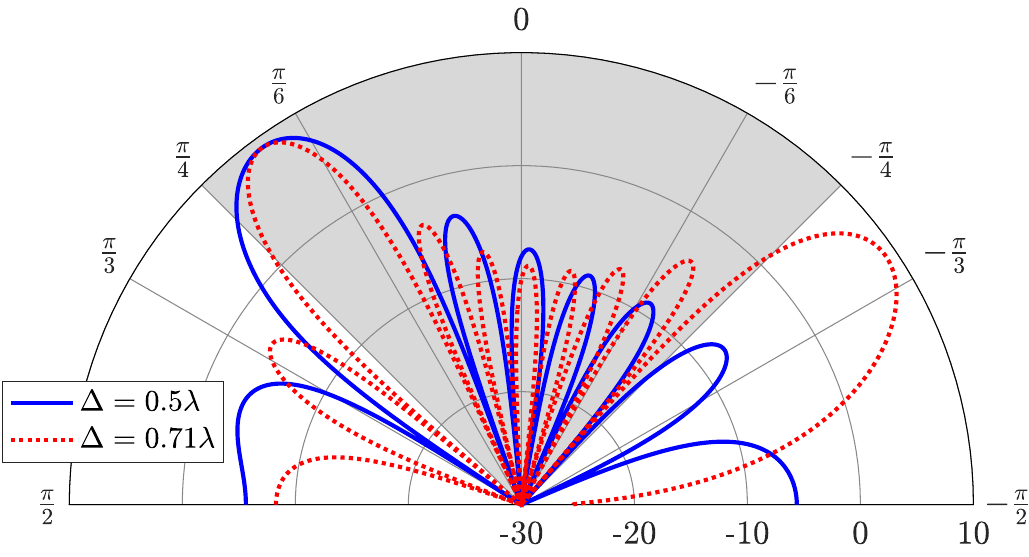}
	 \put(57,44){\footnotesize \rotatebox{-20}{Coverage}}
	 \put(57.5,39){\footnotesize \rotatebox{-20}{region}}
	 \put(71,-3){\footnotesize Array gain [dB]}
	 \put(91,36){\color{black}\vector(-1,-1){5}}
	 \put(88.5,41.5){\footnotesize Grating}
	 \put(91,37){\footnotesize lobe}
\end{overpic}  \vspace{2mm}
                \caption{Beamformed transmission toward angle $\pi/5$.}  
        \end{subfigure} 
        
        \caption{The beampatterns when transmitting in two different directions using a ULA with $8$ antennas. The array gain is independent of the antenna spacing $\Delta$, but the beamwidth shrinks as it increases, resulting in more and larger sidelobes.}
\label{fig_sparser} \vspace{-2mm} 
\end{figure}

The consequences of increasing the antenna spacing beyond $\lambda/2$ are illustrated in Fig.~\ref{fig_sparser} for an eight-antenna ULA with isotropic elements and a spacing of either $\lambda/2$ or $\lambda/ \sqrt{2} \approx 0.71 \lambda$, where the latter is designed for the shaded coverage region.
Fig.~\ref{fig_sparser}(a) shows the beampattern when transmitting toward $0$ radians (broadside direction). We observe that the main lobe retains its array gain but becomes narrower and more sidelobes appear as the antenna spacing increases. 
The narrower main lobe enables us to squeeze $M=8$ different beamformed signals into the shaded angular interval (e.g., toward equally spaced users) so that each peak is at the others' nulls.
With $\lambda/2$-spacing, we can only fit $8/\sqrt{2} \approx 5$ beamformed signals into that interval.
Hence, if we know that the array will only be used for communications within the shaded region, the wider spacing is desirable, as it enables the spatial resolution to be focused on that region.

Since the same amount of power is radiated regardless of the antenna spacing, a narrower main lobe necessarily leads to stronger sidelobes. This can be seen in Fig.~\ref{fig_sparser}(a), but becomes particularly evident when we point the beam towards $+\pi/5$ in Fig.~\ref{fig_sparser}(b).
There is then a sidelobe at $-0.31 \pi$ that is equally strong as the main lobe, leading to substantial interference in an unintended direction. This is known as a \emph{grating lobe} and is often attributed to poor array design; however, in this case, can only appear outside the array's intended coverage region.
Grating lobes are only problematic if we are concerned with interference caused outside the shaded region (due to regulations or coexistence with neighboring systems), or if an array is utilized for a wider coverage region than it was originally designed for. For example, if the array is used for angle-of-arrival estimation, it will be unable to distinguish between signals arriving from $+\pi/5$ and $-0.31 \pi$. This is an instance of the classical \emph{aliasing effect}, but the ambiguity is immediately resolved if we have prior information revealing which alias is correct; in this case, it is always the one in the shaded region.

These properties are only conservatively used in practical systems. The 5G Massive MIMO array in Fig.~\ref{fig_massivemimo}(b) uses $\lambda/2$ spacing horizontally, but $0.7\lambda$ vertically. The latter enhances the spatial resolution when serving users at different elevation angles underneath the BS, at the expense of occasionally sending grating lobes into the sky, where no users are supposed to be.
This is only the first step toward tailoring arrays to the intended coverage region and we discuss more advanced approaches below.

\vspace{-2mm}

\section*{Non-Uniform Sparse Linear Arrays}

The previous section advocated for \emph{sparse arrays} with antenna spacings greater than $\lambda/2$, as a means to shrink the beamwidth when using a fixed number of antennas. It is actually mostly the aperture length---the distance between the outermost antennas---that determines the beamwidth of the main lobe, while the inner antenna arrangement affects the sidelobe patterns~\cite{Skolnik1964a}.
Motivated by this, we will now consider non-uniformly spaced antennas and fine-tune their locations as an additional design dimension for linear arrays.
It remains essential to keep all antenna spacings greater than $\lambda/2$ to avoid mutual coupling, so we call this topic \emph{non-uniform sparse arrays}.

Non-uniform arrays have a long history in localization and sensing applications~\cite{Skolnik1964a}.
The resolution in multi-target detection is determined by the number of unique antenna spacings that exist in the array, considering all ${M \choose 2} = M (M-1) / 2$ antenna pairs. The reason is that each incident planar wave gives rise to a specific apparent wavelength, which can be identified by computing the phase difference between an antenna pair. When there are multiple incident waves, we must have multiple antenna pairs with different separations to distinguish them. ULAs are suboptimal from the redundancy viewpoint, as the same spacings are repeated multiple times.

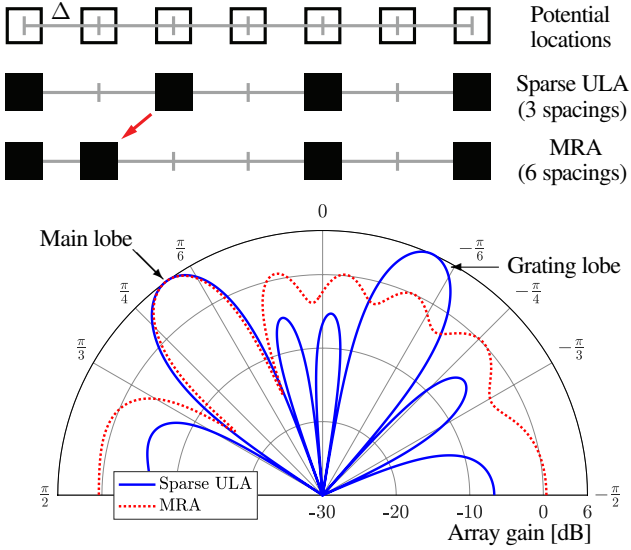
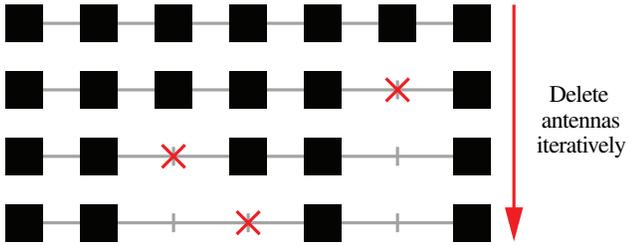
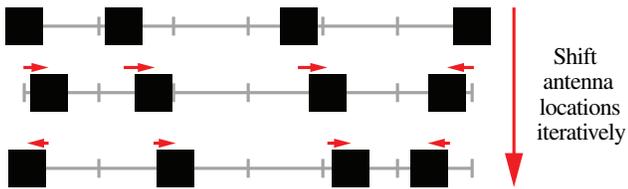
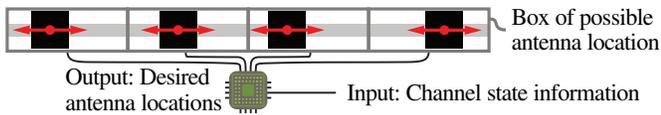
\begin{figure}[t!]
        \centering 
        \begin{subfigure}[b]{\columnwidth} \centering 
	\begin{overpic}[width=.95\columnwidth,tics=10]{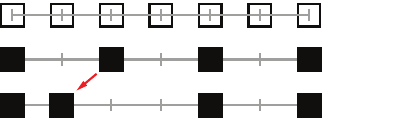} 
	 \put(7.2,26.5){\footnotesize \rotatebox{0}{$\Delta$}}
	 \put(83,26){\footnotesize \rotatebox{0}{Potential}}
	 \put(83,22){\footnotesize \rotatebox{0}{locations}}
	 \put(80.7,15){\footnotesize \rotatebox{0}{Sparse ULA}}
	 \put(81,11){\footnotesize \rotatebox{0}{(3 spacings)}}
	 \put(86,5){\footnotesize \rotatebox{0}{MRA}}
	 \put(81,1){\footnotesize \rotatebox{0}{(6 spacings)}}
 \end{overpic}  \vspace{2mm}
 	\begin{overpic}[width=.9\columnwidth,tics=10]{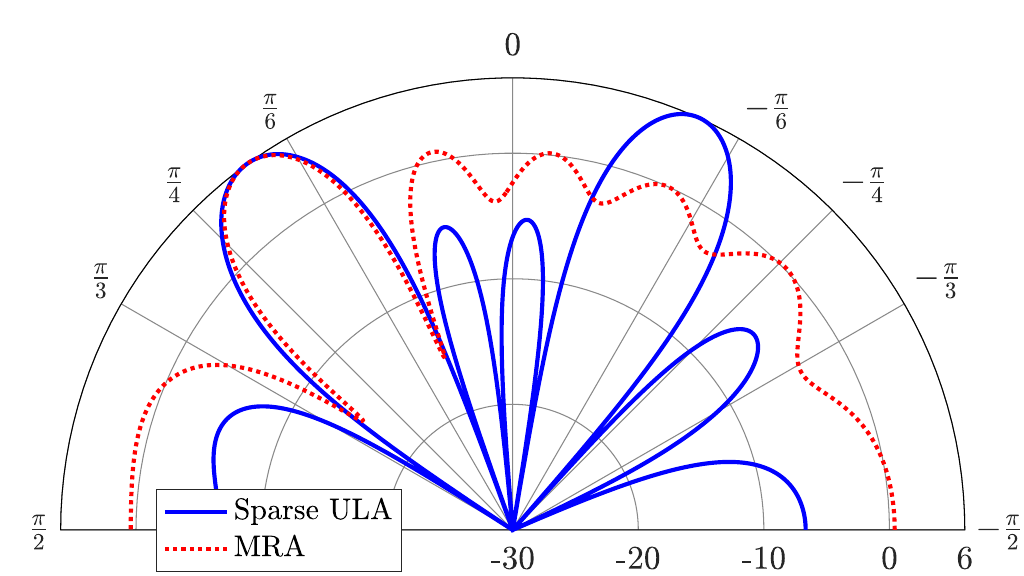}
	 \put(71,-3){\footnotesize Array gain [dB]}
	 \put(3,46){\footnotesize Main lobe}
	 \put(18.5,45.5){\color{black}\vector(1,-1){5}}
	 \put(79.5,42.5){\color{black}\vector(-1,0){8}}
	 \put(81,41.5){\footnotesize Grating lobe}
\end{overpic} \vspace{1mm}
                \caption{Rearranging antennas in a ULA to obtain a minimum-redundancy array that lacks grating lobes and has nearly uniform sidelobes.}   \vspace{+4mm}
        \end{subfigure}    \vspace{+4mm}
        \begin{subfigure}[b]{\columnwidth} \centering
	\begin{overpic}[width=.95\columnwidth,tics=10]{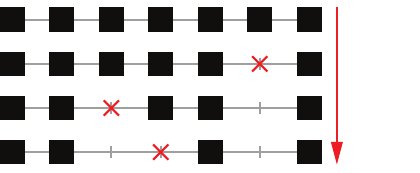}
	 \put(86,24){\footnotesize \rotatebox{0}{Delete}}
	 \put(84.7,20){\footnotesize \rotatebox{0}{antennas}}
	 \put(84,16){\footnotesize \rotatebox{0}{iteratively}}
\end{overpic}  
                \caption{Designing a sparse array through thinning of a dense array.}  \vspace{+2mm}
        \end{subfigure} 
        \begin{subfigure}[b]{\columnwidth} \centering
	\begin{overpic}[width=.95\columnwidth,tics=10]{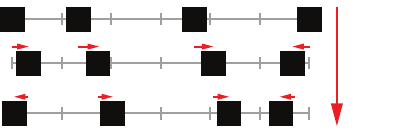}
	 \put(86.6,21){\footnotesize \rotatebox{0}{Shift}}
	 \put(85,17){\footnotesize \rotatebox{0}{antenna}}
	 \put(84.4,13){\footnotesize \rotatebox{0}{locations}}
	 \put(84,9){\footnotesize \rotatebox{0}{iteratively}}
\end{overpic}  
                \caption{Designing a PIA by adjusting the antenna locations.}  \vspace{+3mm}
        \end{subfigure} 
        \begin{subfigure}[b]{\columnwidth} \centering
	\begin{overpic}[width=.95\columnwidth,tics=10]{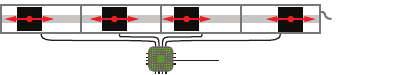}
	 \put(80,14){\footnotesize \rotatebox{0}{Box of possible}}
	 \put(80,10){\footnotesize \rotatebox{0}{antenna location}}
	 \put(54,2){\footnotesize \rotatebox{0}{Input: Channel state information}}
	 \put(9.5,4.5){\footnotesize \rotatebox{0}{Output: Desired}}
	 \put(9.5,0.5){\footnotesize \rotatebox{0}{antenna locations}}
\end{overpic}  
                \caption{Intelligent array with real-time movable antennas.}  \vspace{+2mm}
        \end{subfigure} 
        \caption{Non-uniform arrays can follow different design principles, using different features and optimizing different metrics.}
\label{fig_nonuniform} \vspace{-2mm}
\end{figure}

This fact has given rise to alternative \emph{minimum-redundancy array (MRA)} designs~\cite{Moffet1968a}, which build on the \emph{Golomb ruler} concept: a ruler with a nonuniform set of marks at integer locations such that no two inter-mark distances are equal.
Similarly, an MRA is designed so that all antenna pairs have different spacings but are multiples of a base interval $\Delta$ (selected based on the sampling theorem for the given coverage region). We exemplify such array design in Fig.~\ref{fig_nonuniform}(a), where the aperture length is $6\Delta$, so there are seven potential antenna locations, but we only deploy four antennas.
A sparse ULA with $2\Delta$-spacing introduces redundancy: there are three antenna pairs with $2\Delta$-spacing, two with $4\Delta$, and one with $6\Delta$. By moving one of the antennas as shown in Fig.~\ref{fig_nonuniform}(a), we obtain an array where each of the six spacings $\Delta,2\Delta,\ldots,6\Delta$ appears only once. This is a perfect MRA since all of the ${4 \choose 2}=6$ spacings are unique.

The beampatterns achieved when these arrays transmit towards $\pi/5$ are also shown in the figure. The main lobes are almost identical, but when we moved one antenna to achieve an MRA, we drastically changed the sidelobes. There is no grating lobe anymore which is important for unambigious target detection, but instead the interference is spread more uniformly since the sidelobes are stronger and wider.

A perfect $M$-antenna MRA has $M (M-1)/2$ unique spacings between the antenna pairs and can utilize them to distinguish an equal number of targets. It is extraordinary that an array can distinguish many more targets than the number of antennas, but the number remains limited by the aperture length---we simply ``delete'' antennas within the array when they provide redundant information.
This results in a sparse array with a narrow beamwidth, while the non-uniformity prevents grating lobes. Unfortunately, target detection with MRAs is generally computationally complex when the array is unstructured. Hence, it is common to introduce some structure and redundancy to aid algorithmic development and make it easier to scale the array designs to different antenna numbers. \emph{Nested arrays}~\cite{Pal2010a} and co-prime arrays represent special cases where combinations of ULAs with different spacings are used to achieve reduced-complexity designs.

MRAs are suitable for channel estimation in communication systems operating in sparse propagation environments because they enable the estimation of individual multipaths using localization-like methods. However, MRAs were originally not designed for effective data transmission. Even if we can detect more than $M$ targets, we cannot communicate with more than $M$ users simultaneously since $M$ is the largest number of linearly independent $M$-dimensional beamforming vectors. However, it gives inspiration for designing non-uniform sparse arrays tailored for communications.
Fig.~\ref{fig_nonuniform}(b) illustrates the concept of \emph{thinning}~\cite{Skolnik1964a}, where we begin with a dense ULA having the intended aperture length but an abundance of antennas. We then delete antennas iteratively until the desired number is reached (e.g., $M=4$), which results in a non-uniform sparse array. 
The deletion is done greedily to maintain a high performance value, such as the average sum rate in the coverage region. 

Fig.~\ref{fig_nonuniform}(c) shows the alternative \emph{pre-optimized irregular array (PIA)} approach~\cite{Irshad2025a}, which starts with exactly $M$ antennas and then fine-tunes their locations. 
The adjustments of the antenna locations can be implemented to maximize the average performance (e.g., sum rate) using particle swarm optimization. This is more flexible than thinning since we are not limited to placing antennas at $\Delta$-spaced grid points. The pre-optimization can be implemented by building a digital twin of the coverage region, where large amounts of synthetic channel data can be gathered for arbitrary antenna locations. This is akin to how telecom operators currently use ray-tracing methods to determine the desired BS locations and downtilt.

\section*{Intelligent Arrays with Movable Antennas}

The PIA consists of antennas at fixed locations identified by moving them around in the simulator \emph{before} deployment, thereby tailoring the array to the geographical properties of the presumed coverage region. However, if we could shift the antenna locations \emph{after} deployment, we could also adapt the array to the small-scale fading properties of the currently scheduled users. Mechanisms for reconfiguring the beampatterns of antenna arrays have been developed for decades. In addition to using traditional phase-shifter networks for beamforming, the concept of mechanically \emph{movable antennas (MAs)} was introduced in~\cite{Pan2007a} describing hardware capable of physical antenna movements. A vision for utilizing these features to create intelligent arrays for MIMO communications was presented in~\cite{Bjornson2019c} and has led to a vast literature studying arrays of MAs~\cite{Zhu2025a}.

Fig.~\ref{fig_nonuniform}(d) illustrates a linear array consisting of MAs, which are connected to a control unit that tunes the antenna locations based on CSI and other characteristics to enhance the system performance. Each antenna can be moved within a predefined box; for example, by using rails to move metal elements or reconfiguring an electrolytic fluid. One can also take an antenna-selection-inspired approach where each of the $M$ transceiver chains has a set of fixed radiating elements that it can choose between using a switch.
We are not concerned with the specific implementation details in this paper (see e.g.,~\cite{Zhu2025a} for a recent review), but with what an intelligent array could potentially achieve in future systems. The vision with MAs is to alleviate the need for excessive transceiver hardware by integrating intelligence directly into the array.

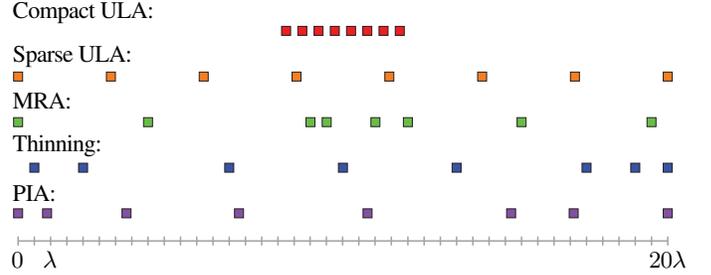
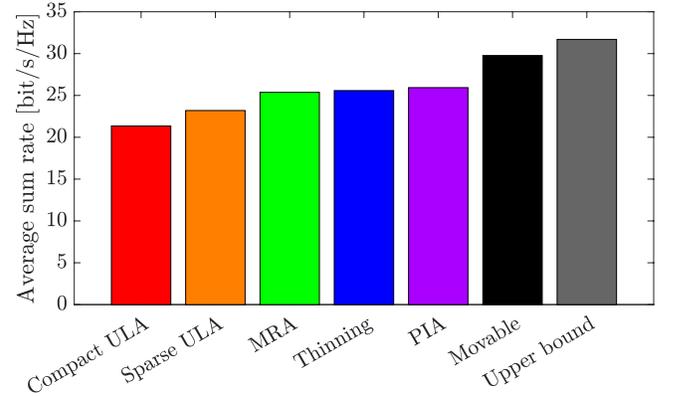
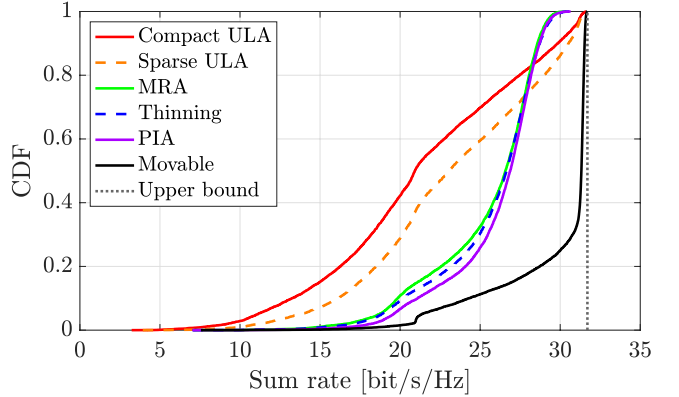
\begin{figure}[t!]
        \centering 
        \begin{subfigure}[b]{\columnwidth} \centering 
	\begin{overpic}[width=\columnwidth,tics=10]{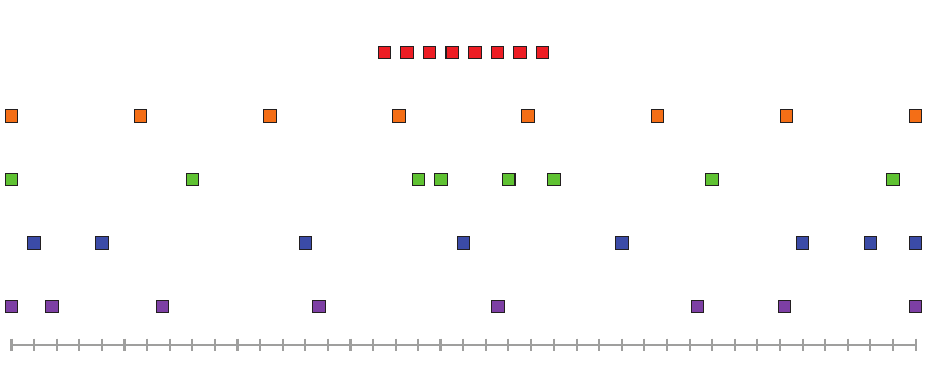} 
	 \put(0.5,37.5){\footnotesize Compact ULA:}
	 \put(0.5,31){\footnotesize Sparse ULA:}
	 \put(0.5,24){\footnotesize MRA:}
	 \put(0.5,17.5){\footnotesize Thinning:}
	 \put(0.5,10){\footnotesize PIA:}
	 \put(0.3,0){\footnotesize $0$}
	 \put(5,0){\footnotesize $\lambda$}
	 \put(96,0){\footnotesize $20\lambda$}
 \end{overpic}  
                \caption{Illustration of the five fixed antenna arrays.}   \vspace{+3mm}
        \end{subfigure} 
        \begin{subfigure}[b]{\columnwidth} \centering
	\begin{overpic}[width=\columnwidth,tics=10]{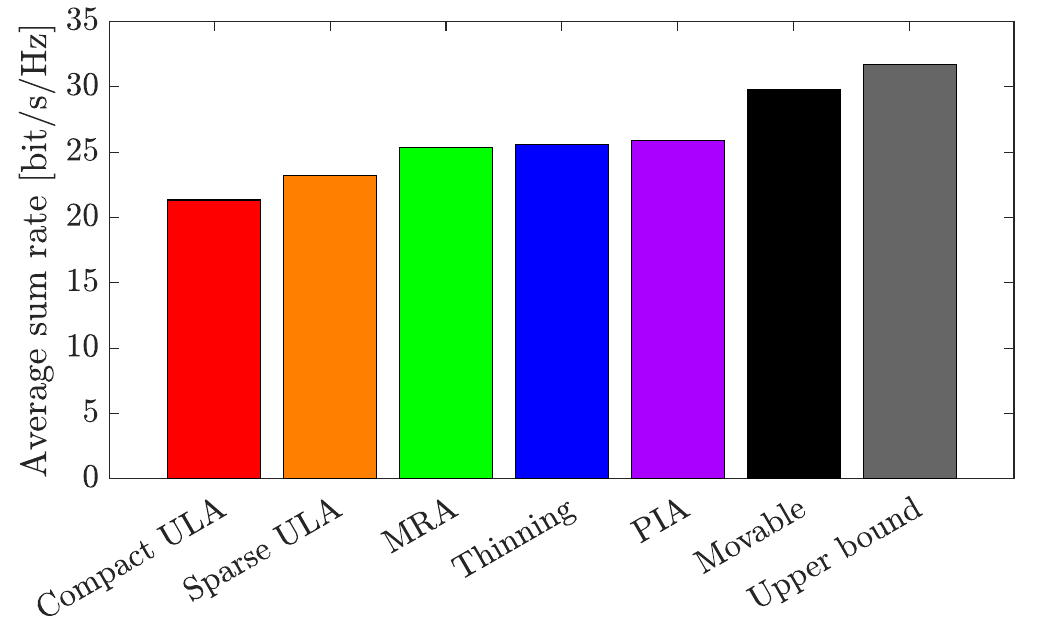}
\end{overpic}  
                \caption{The average sum rate achieved for different user locations.}  \vspace{+4mm}
        \end{subfigure} 
        \begin{subfigure}[b]{\columnwidth} \centering
	\begin{overpic}[width=\columnwidth,tics=10]{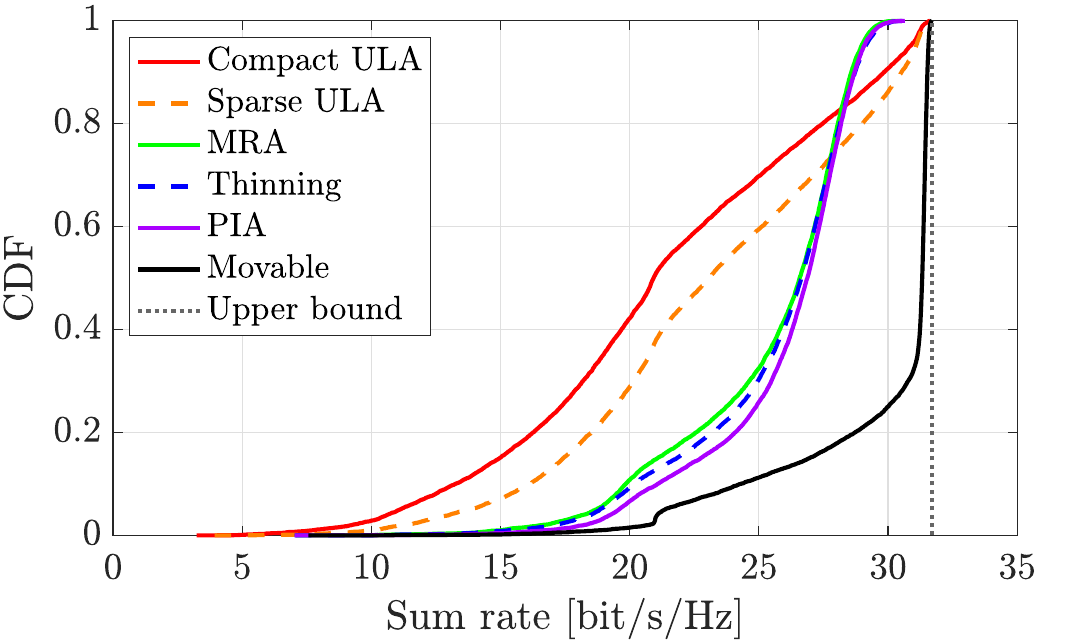}
\end{overpic}  
                \caption{CDF of the sum rate achieved for varying user locations.}  \vspace{+2mm}
        \end{subfigure} 
        \caption{A comparison of the uplink sum rate achieved with six linear arrays in a LOS scenario with random user angles. The ULAs are outperformed by the non-uniform arrays, but the array with MAs provides the highest rates.} 
\label{fig_cdf_example} \vspace{-3mm}
\end{figure}

We will now analyze the communication performance in an uplink simulation scenario, where $K=5$ users are located in random angular directions, uniformly distributed between $\pi/4$ and $-\pi/4$. We consider LOS channels with $10$\,dB SNR per antenna. The cumulative distribution function (CDF) of the sum rate for different user locations, as well as the average sum rate, will be compared for different antenna arrays. The BS has an array of $M=8$ antennas that can be deployed within a linear interval of length $20\lambda$.
We consider MAs, where the array is fine-tuned for each realization of the user locations, and compare it with the five fixed antenna arrays illustrated in Fig.~\ref{fig_cdf_example}(a). There is a compact ULA with $\lambda/2$-spacing and a sparse ULA that spans the maximum aperture length.
There are also three non-uniform sparse arrays: an MRA from~\cite{Moffet1968a}, an array obtained by thinning a ULA that originally had $41$ elements, and a PIA optimized following the approach from~\cite{Irshad2025a}. 

The average uplink sum rate achieved with RZF beamforming and perfect CSI is shown in Fig.~\ref{fig_cdf_example}(b). The sparse ULA gives higher rates than the compact ULA because it has narrower main lobes, which makes it easier to serve closely spaced users. The three non-uniform sparse arrays provide even higher rates, as they alleviate the grating lobes that the sparse ULA suffers from, which makes it easier to deal with remaining inter-user interference using beamforming. The array with MAs provides even higher rates because it can tailor the sidelobe pattern so the nulls point toward unintended users. Despite this excellent ability, a $7\%$ gap remains to the upper bound where all interference is neglected.

To dig deeper into these results, we show the CDF of the sum rate (with respect to different user locations) in Fig.~\ref{fig_cdf_example}(c).
The ULAs exhibit huge performance variations depending on the users' relative locations; the upper bound is achieved when all users happen to be widely spaced, but the sum rate collapses when the array struggles to separate closely spaced users. The non-uniform arrays provide more stable performance, with higher guaranteed rates, but also a wider gap to the upper bound. This is a consequence of optimizing for the average case, which will smear out sidelobe interference over many angles (as shown in Fig.~\ref{fig_nonuniform}(a)).
The MAs' performance is close to the interference-free upper bound for the majority of user realizations; however, it is not a silver bullet, as the main-lobe width is barely changed through antenna movements, but mostly the sidelobe pattern. Hence, if two users are too closely spaced, their mutual interference cannot be alleviated by adjusting the antenna locations.

\section*{Optimization of Non-Uniform Planar Arrays}

The beams transmitted from horizontal linear arrays have a limited horizontal beamwidth, but resemble orange slices when observed in 3D.
To distinguish between users both horizontally and vertically, outdoor 5G BSs are equipped with planar antenna arrays that create cone-shaped beams.
UPAs with a rectangular grid of antennas, as in Fig.~\ref{fig_massivemimo}(b), are common in practice. However, the spatial samples taken by such arrays contain much redundancy because each row observes the waves' horizontal dimension identically, while each column observes the vertical dimension identically.

One can achieve similar spatial resolution but with vastly fewer antennas by deploying a combination of two linear arrays, in the shape of an L, T, or +.
Inspired by this fact, the literature contains a plethora of non-uniform array designs for localization/sensing purposes. 
The recent review~\cite{Aboumahmoud2021a} describes as many as 38 such array constructions, combining multiple linear arrays, nesting of UPAs with different spacings, or resembling boxes with holes in the middle.

There is no universally optimal planar array shape for localization, but it depends on the preferred balance between multiple performance metrics (e.g., horizontal/vertical angular resolution) and physical constraints (e.g., size and power).
For wireless communication purposes, we can use the average sum rate as the metric and exploit thinning or PIA methods to design an array tailored for a specific BS site. To demonstrate the potential benefits, we consider a refined 3D version of the previous simulation scenario. The BS is equipped with $M=16$ antennas and serves $K=10$ users randomly distributed on the ground in a $120^\circ$ sector at distances from $100$ to $200$ meters, leading to uplink SNRs of $7$--$22$\,dB per antenna. We consider a wideband Rician channel model with six clusters, a k-factor of $8$\,dB, and a carrier frequency of $3$\,GHz. We compute average sum rates over fifty $15$\,kHz-subcarriers to capture the impact of frequency-selective fading. 

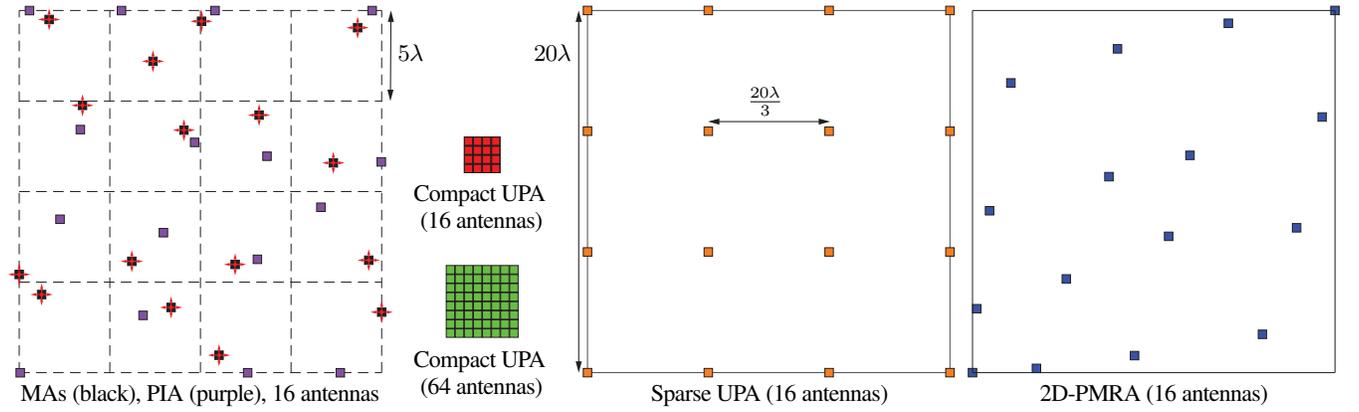
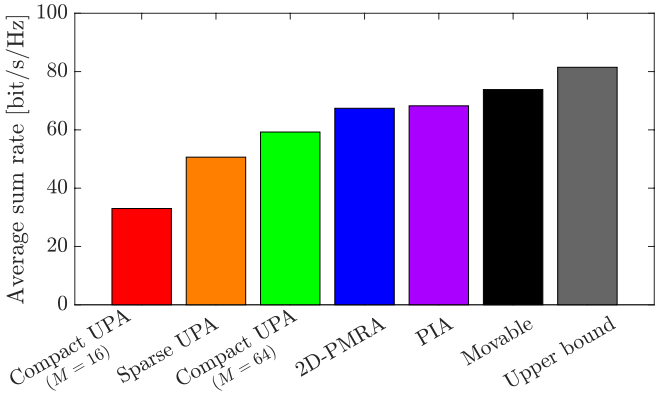
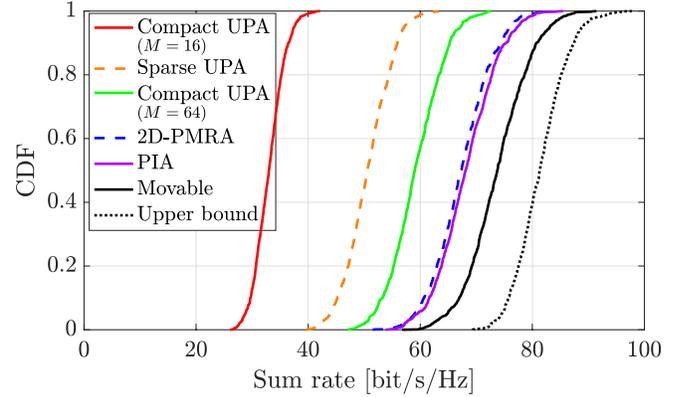
\begin{figure*}[t]
        \centering 
        \begin{subfigure}[b]{2\columnwidth} \centering 
	\begin{overpic}[width=\columnwidth,tics=10]{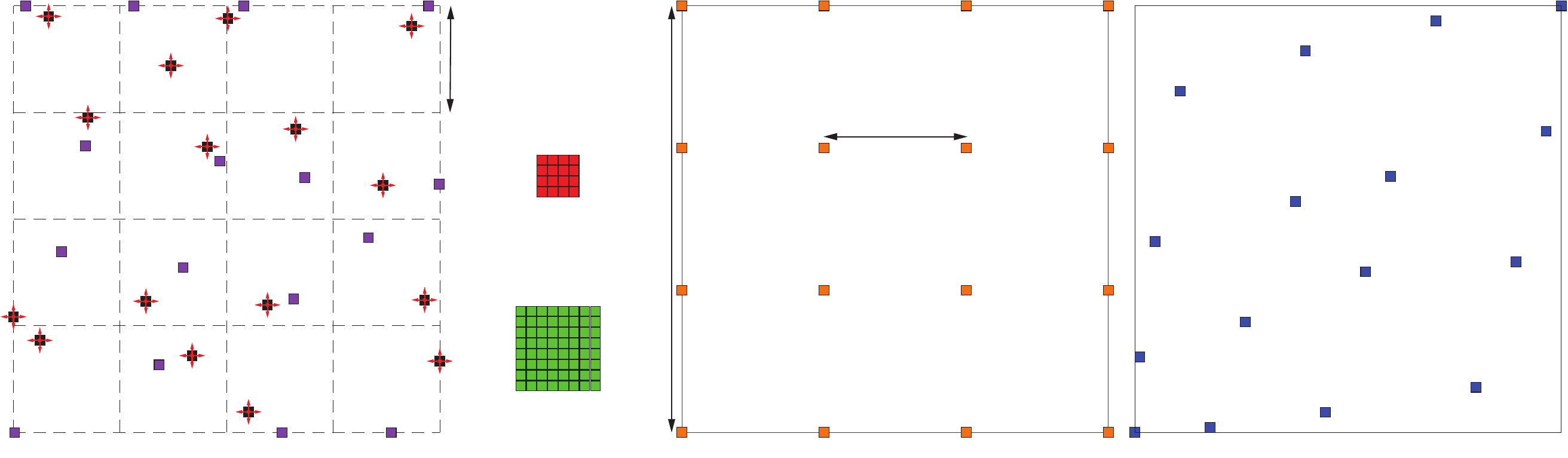} 
	 \put(30.5,14.5){\footnotesize Compact UPA}
	 \put(31,12.5){\footnotesize (16 antennas)}
	 \put(30.5,2){\footnotesize Compact UPA}
	 \put(31,0){\footnotesize (64 antennas)}
	 \put(48.3,-0.5){\footnotesize Sparse UPA (16 antennas)}
	 \put(77.5,-0.5){\footnotesize 2D-PMRA (16 antennas)}
	 \put(1,-0.5){\footnotesize MAs (black), PIA (purple), 16 antennas}
	 \put(29.3,25){\footnotesize $5\lambda$}
	 \put(55.5,21.5){\footnotesize $\frac{20\lambda}{3}$}
	 \put(39.5,25){\footnotesize $20\lambda$}
 \end{overpic}    \vspace{-1mm}
                \caption{Illustration of five fixed planar antenna arrays and the realization of the MAs for one set of user locations.}   \vspace{+2mm}
        \end{subfigure} \\
        \begin{subfigure}[b]{\columnwidth} \centering
	\begin{overpic}[width=\columnwidth,tics=10]{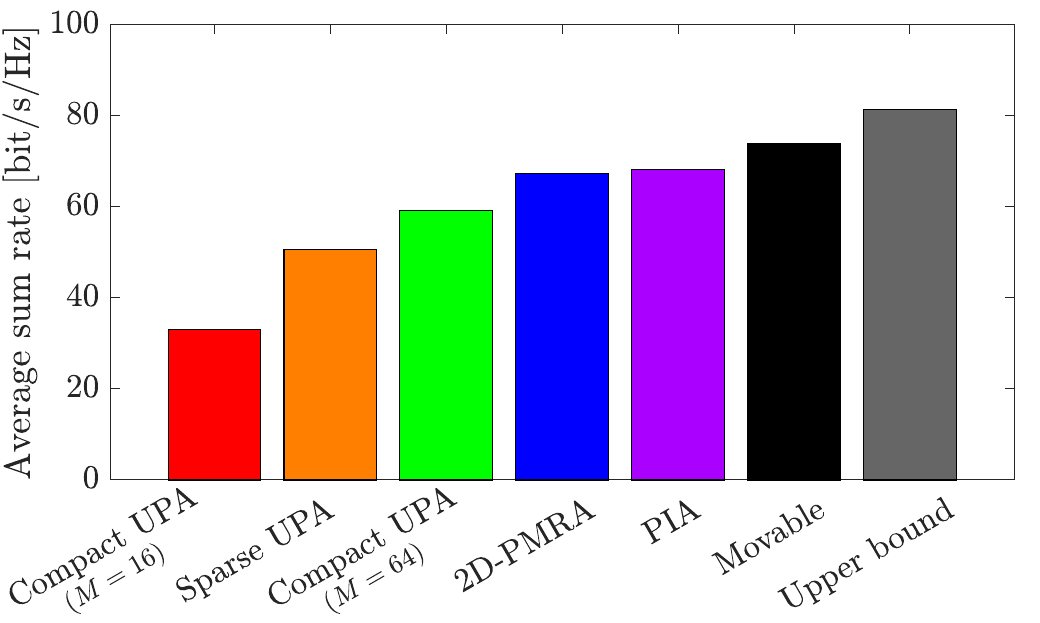}
\end{overpic}  
                \caption{The average sum rate achieved over different user locations.}
        \end{subfigure} 
        \begin{subfigure}[b]{\columnwidth} \centering
	\begin{overpic}[width=\columnwidth,tics=10]{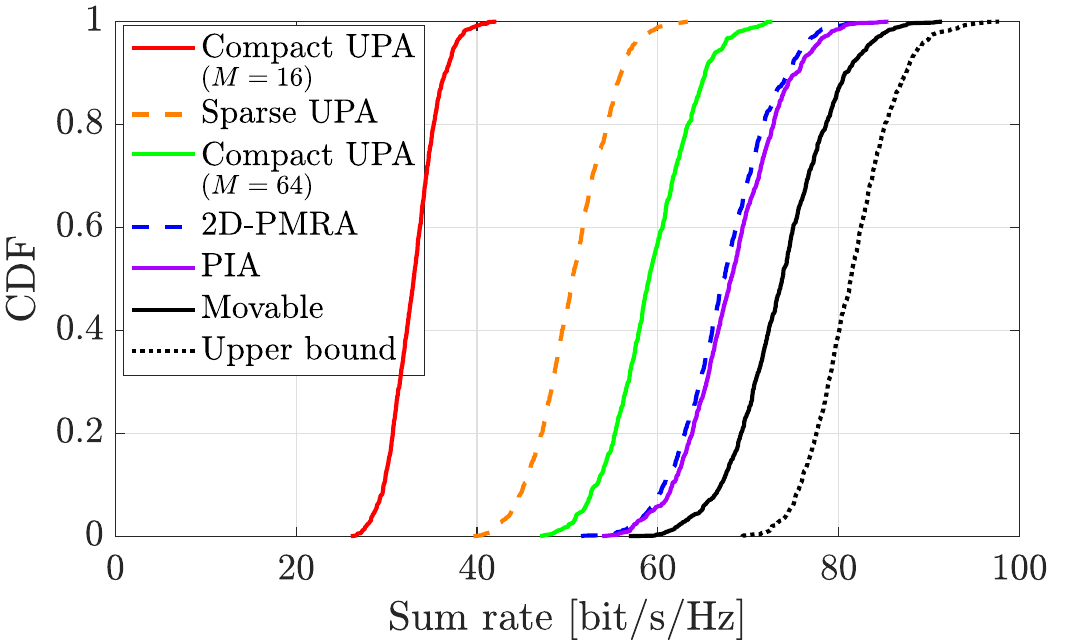}
\end{overpic}  
                \caption{CDF of the sum rate achieved for varying user locations.}
        \end{subfigure} 
        \caption{A comparison of the uplink sum rate achieved with six different planar arrays in a Rician channel with a dominant LOS component. All arrays except one have 16 antennas. The UPAs (even the one with 64 antennas) are outperformed by the fixed non-uniform PIA and 2D-PMRA arrays with only 16 antennas each, but the array with MAs provides the highest rates.} 
\label{fig_2D_example} \vspace{-3mm}
\end{figure*}

Fig.~\ref{fig_2D_example}(a) illustrates six different planar BS arrays that all fit on a $20\lambda \times 20\lambda$ surface facing the coverage region. 
The baselines consist of two square UPAs with a $\lambda/2$-spacing and either $16$ or $64$ antennas. The remaining arrays are sparse and have $16$ antennas. We consider a sparse UPA with the maximum antenna spacing and a non-uniform array that we call 2D-permuted MRA (2D-PMRA). The latter was created from a $16$-antenna linear MRA by using its positions as both horizontal and vertical coordinates, but with a permutation so the array resembles a skewed UPA.
Finally, we consider a PIA and an array with MAs, both optimized under the constraint that each antenna can only be placed within a designated $5\lambda \times 5 \lambda$ region. The optimization was implemented as in~\cite{Irshad2025a}, which means that the PIA optimizes the average sum rate over $1000$ user drops, while the MAs are reoptimized for each user drop.

The average uplink sum rates are provided in Fig.~\ref{fig_2D_example}(b) and show much larger differences than in the linear array scenario in Fig.~\ref{fig_cdf_example}: MAs increase the sum rate by $124$\%, PIA by $107$\%, and the sparse UPA by $53$\% compared to the 16-antenna compact UPA.
In fact, we even outperform the 64-antenna compact UPA using 16-antenna designs based on either MAs, PIA, or 2D-PMRA.
This is the result alluded to in the paper title: \emph{we can alleviate the antenna abundance in 5G Massive MIMO by using sparse arrays with fewer, but more strategically positioned, antennas}. Since the 2D-PMRA performs nearly as well as the PIA, the important thing is to use a judiciously designed non-uniform array, rather than directly optimizing it for communications.

CDF curves showing the sum rate variations over different user drops are provided in Fig.~\ref{fig_2D_example}(c). All the curves have similar shapes, which was not the case in the linear array scenario. The reason is that we consider a more realistic propagation environment with non-LOS paths, which helps the BS to distinguish between users that have similar LOS paths. Another key difference is that the array with MAs never reaches the interference-free upper bound, which is attributed to the frequency-selective fading; the antenna locations cannot be optimized to provide fully orthogonal user channels on all subcarriers, but they can still make the LOS-path portion of the channel vectors orthogonal.

\vspace{-2mm}

\section*{Maintaining Link Budget using Fewer Antennas}

If we reduce the number of antennas, $M$, in future systems, we must be mindful not to degrade the link budget per user.
The SNR is proportional to $M \cdot G$, where $G$ is the antenna gain. Hence, we must increase the antenna gains when decreasing the number of antennas.
This was implicitly done in the simulations reported in Fig.~\ref{fig_2D_example}, where $M \cdot G$ was the same for the $16$- and $64$-antenna arrays.
One way to fine-tune $G$ is to let each antenna consist of a subarray of multiple radiating elements, as illustrated in Fig.~\ref{fig_massivemimo}(b), where each antenna comprises two vertically-stacked elements.
There exist 5G Massive MIMO products with $64$ or $32$ antennas but the same total number of radiating elements~\cite{Bjornson2019c}, so that $M \cdot G$ is the same. 
A similar approach can be taken when building non-uniform Gigantic MIMO arrays.

Alternatively, a transmissive reconfigurable intelligent surface (RIS) can be placed in front of a non-uniform sparse array with low-gain antennas, and a metallic enclosure can guide electromagnetic signals between them to create a quasi-horn antenna system~\cite{Demir2024a}. Such an RIS-integrated array has an effective aperture determined by the RIS' size, rather than the number of antennas. Hence, it collects more energy in the uplink and focuses the energy efficiently in the downlink, essentially making the link budget independent of the number of antennas.
Moreover, the RIS's configuration can be modified slowly based on user statistics, or dynamically based on CSI, to point the directivity towards where users are.

\vspace{-2mm}

\section*{Conclusions}

Non-uniform planar arrays are superior to UPAs in multiuser MIMO communications. This fact can be leveraged to design future Gigantic MIMO systems with a smaller antenna-user-ratio than in 5G. The highest sum rate is achieved with intelligent arrays of real-time movable antennas, but one can also come very far in the 6G era with judiciously designed, yet fixed, non-uniform arrays.

There are numerous opportunities for future research on this topic, including the tailored design of signal processing schemes (e.g., channel estimation and codebook design) and transceiver hardware for non-uniform arrays (e.g., RIS-integrated arrays and movable antennas), as well as support for multiple applications (e.g., integrated sensing and communications). Together, these directions point towards the need for co-design between array design and services.

\vspace{-2mm}

\balance
\bibliographystyle{IEEEtran}
\bibliography{IEEEabrv,refs}

\end{document}